# Crowd-Funded Earthquake Early-Warning System

Hudson Kaleb Dy
Research Associate
Walnut Valley Research Institute

Hsi-Jen James Yeh, Ph.D.
Associate Professor
Department of Engineering & Computer Science
Azusa Pacific University

*Abstract*—Earthquake early warning systems has been proven to save countless lives in Japan, Mexico, and Chile, where earthquake warnings are often broadcast live on TV up to a minute before residents experience shaking.

Unfortunately, traditional early warning systems require extensive capital investment. The high cost of traditional earthquake early-warning systems and limited budgets prevent earthquake-prone developing countries like the Philippines, Indonesia, Afghanistan, India, Burma, Ghana, Nigeria, Columbia, Venezuela, and Bolivia from building traditional earthquake warning systems.

This project describes repurposing old Android smartphones into affordable dedicated seismometers to detect tremors. These smartphones have become disposable items and are continuously "upgraded" and replaced. Yet every one of these devices includes everything needed to act as a dedicated seismometer: Wi-Fi capability, GPS, and an accelerometer.

The software developed for this project converts these smartphones into dedicated seismometers and uses existing web technologies for telemetry services. This system would also trigger alerts to all devices that has the software installed whenever a tremor is detected, effectively making each seismic detection station double as an earthquake early-warning alarm.

A large network of these seismic detection stations will effectively create an affordable earthquake early-warning system that can be rapidly implemented at an extremely low cost. It would provide developing nations an affordable life-saving alternative to expensive traditional earthquake early-warning systems. This solution is cheap, keeps old smartphones from landfills, and will save lives. (*Abstract*)

## I. INTRODUCTION

When an earthquake occurs, several types of seismic waves are generated from the epicenter. These include less destructive but fast-moving P-wave (primary waves) [1] as well as more destructive but slower moving S-waves (secondary waves).

Earthquake early-warning systems work by distributing seismic detection station in a widespread area[1]. When an earthquake occurs, the detection stations closest to the epicenter will detect the fast-moving P-waves to warn of the impending destructive S-waves. Although residents located closer to the epicenter will have little if any advance warning, those located further will have critical seconds, and sometimes minutes, to brace for the shaking. It would give people who live in seismically vulnerable structures or in tsunami prone areas extra time to move to safter areas.

Earthquake early warning systems has been proven to save countless lives in Japan, Mexico, and Chile, where earthquake warnings are often broadcast live on TV up to a minute before residents experience shaking.[5][6][15]

Unfortunately, these systems are notoriously absent in developing nations because of the high cost of traditional earthquake early-warning systems. [12][14][16]With one of the highest GDP per capita in the world, it took until May 2021 for the US west coast to finally have a working system even with a big earthquake long overdue.[10][11] This system is called the USGS ShakeAlert system which cost USD 39.4 million to implement the network of seismometers, plus another USD 20.5 million for telemetry[2][3][4]. Annual maintenance costs of the network of seismometers is USD 28.6 million per year, plus USD 9.8 million for the telemetry system. The total implementation cost to operate the USGS ShakeAlert system for the first year is USD 98.3 million (USD 59.9 million + USD 38.4 million.)

This project is an attempt to bridge the gap and provide a crowd-funded affordable solution for earthquake-prone developing countries.

To reduce cost, this project repurposes old Android smartphones into affordable dedicated seismometers to detect tremors. These smartphones have become disposable items and are continuously "upgraded" and replaced. USA disposes 151 million cellphones per year. [17] Yet every one of these devices includes everything needed to act as a dedicated seismometer: Wi-Fi capability, GPS, and an accelerometer. A secondary benefit of this project is the ability to reuse these old cellphones and prevent them from going into landfills.

## II. SEISMIC DETECTION STATIONS

The current standard of building a traditional earthquake early-warning system involves setting up dedicated seismic detection stations. It would involve leasing space, providing power, and building a telemetry system for the seismic detection station to communicate with the rest of the system. A central server that processes these signals from each remote detection station is then used to validate if an earthquake occurs or if it is a false signal.

This project leverages existing WiFi and internet connectivity in most homes to eliminate the telemetry costs. In the case of the USGS ShakeAlert system, this cost $20,000,000 for the buildout, plus $9,800,000 annual expenses. Our system would be mounted on existing load bearing walls of homes and other existing buildings with double-sided tape using a crowd-funded approach. This would also eliminate construction costs and land acquisition costs. In the case of USGS ShakeAlert system, this was $39,000,000 for the initial buildout, plus $28,600,000 annual expenses.

III. COMPONENTS

The crowd-funded earthquake-early warning system consists of the following elements:

1) An Android application software specifically created for this project. This is an Android Java application written in Android Studio that will convert the phones into dedicated seismometers when it is installed in an Android smartphone. These devices would be permanently plugged into a power outlet using the original device charger and mounted on a load-bearing wall of the building using double-sided tape.

2) A server component that comprises a Linux Apache web server and several server-side PHP scripts.
3) A MySQL server that is accessible from the web server. This stores all reported shaking events, the time of the event, and the GPS location. A unique ID will be automatically generated and maintained for each device.

By having a large network of old Android smartphones running the software application developed for this project and distributed geographically in an earthquake prone region, this system should work and operate similarly to more expensive traditional earthquake early-warning systems.

*A. Android Smartphone*

The Smartphones will be required to connect to an existing WiFi connection and plugged in to the wall outlet using the original device charger. This will ensure that the device will stay continuously powered. It will also be attached to a load-bearing wall using double-sided tape. Lastly, the Android Software Application developed for this project shall be installed into the device.

*B. Android Software Application*

When installed in an Android smartphone, the application will detect ground shaking by monitoring the accelerometers in the smartphones. It will also gather the device's latitutde and longitude (physical location) using the device's

GPS. Tremors detected will be reported to the server by calling the PHP scripts. The reported data will include information such as the amplitude of the shaking detected, the time and the physical GPS location of the device.

*C. PHP Server-Side Scripting*

Upon receipt of a reported tremor, the server will store the amplitude, time, and GPS location on the MySQL database. It will then proceed to perform a simple validation procedure to determine if the shaking was the result of a tremor or just a localized anomaly (like passing trucks causing a vibration). This validation is done by looking for nearby devices and checking if they reported a similar tremor. If the reported tremor passes validation, the server-side script will send out a notification using Google's Firebase Cloud Messaging Notifications to all devices with the same application installed to warn of a pending earthquake.

*D. Google FCM Notifications*

Google FCM Notifications are used to notify users of a pending large tremor. While users closer to the epicenter will not receive advance notifications, users located further away will benefit by getting advance warning. Google FCM ensures is free and ensures that 95% of users will receive the notification within 250ms and has very low bandwidth requirements. [17]

*E. MySQL Server*

The database is used to quickly verify if nearby stations has detected similar tremors without maintaining dedicated connections to individual Android devices. This is crucial to allowing the system to scale. It also allows us to further tune the system by increasing or decreasing the threshold required, whether it is with regards to the intensity of shaking (amplitude of detected shaking in accelerometer) or the required number of stations that has detecting shaking, before raising an alarm of an impending tremor to all users.

IV. METHODS AND RESULTS

The prototype that was built is fully functional and has demonstrated to work well when tested with 3 devices. For our initial test, because we are lacking sufficient test devices, the validation procedure has been automatically disabled until the number of devices with our software installed hit 50 units.

Because the built-in accelerometers in the cellphones are incredibly sensitive and was able to pick up a lot of environmental shaking such as when large trucks pass by the building structure. Initial version of the program reported all shaking to the server but resulted in many false positive tremor reports. A modified version of the application has filtered this small environmental background activity and will only report moderate or severe shaking.

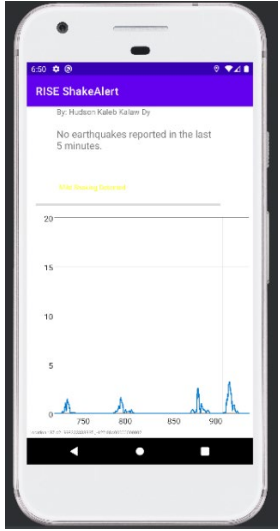

Figure 1: Screenshot of minor shaking detected by device. These are residual environmental shaking and are not reported to the server. No Earthquake warning received from the server in the last 5 min.

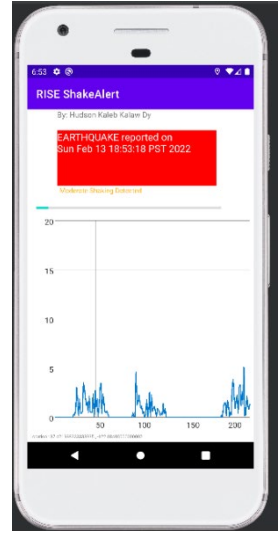

Figure 3: Moderate shaking detected by device which are reported to the server. This is a screenshot after the shaking report was sent to the server and the earthquake warning received back from the server. An audible alarm is also triggered on every device.

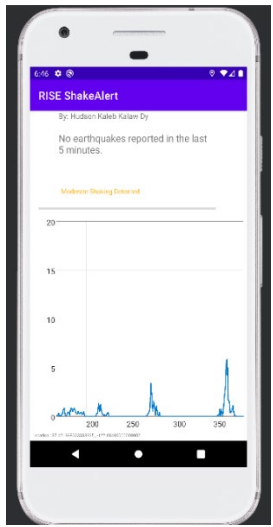

Figure 2: Moderate shaking detected by device which are reported to the server. This shaking is sent to the server, but this screenshot was taken right before the earthquake warning has been received back from the server.

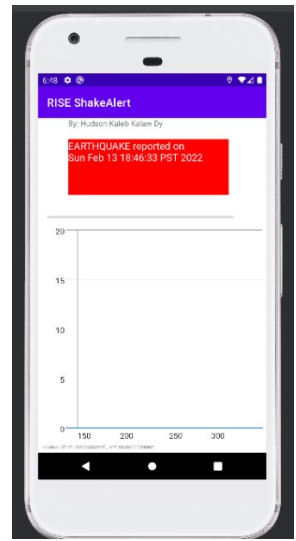

Figure 4: Screenshot of another device showing no shaking detected locally, but an Earthquake warning was received from the server which is shown on screen. An audible alarm is also triggered on every device.

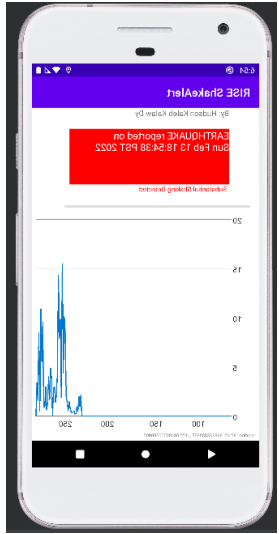

Figure 5: Extreme shaking detected locally and sent to the server. This is a screenshot after the shaking report was sent to the server and the earthquake warning received back from the server. An audible alarm is also triggered on every device.

## V. DISCUSSION

All software has been developed with the completed user interface and the system fully functional. There amount of shaking are roughly differentiated and grouped according to "Minor Shaking", "Moderate Shaking" and "Substantial Shaking." Although these are sufficient to provide warning alerts of a pending earthquake, improvements could be made to be more indicative of the magnitude of the earthquake like traditional earthquake early-warning systems. In order to improve the system to do so, additional work to calibrate different types of phones and the sensitivity of the accelerometers would need to be completed.

The system would also become more effective as the number of dedicated seismometers (repurposed Android cellphones with the software installed) grow. This would need to occur to make the system and viable. Once the installed user base grows sufficiently, this system could conceivably be in TV and radio broadcast stations to stream warnings to the public at large.

Once there are enough dedicated seismic detection stations in place, another version of this software could also be created that only receives warning alarms to send the alert to everyone on their cellphones. This system could also be integrated into the public emergency alert systems to broadcast warnings and alarms to people without cellphones. Making the system crowd-funded should allow the system to rapidly increase in numbers and its ability to sufficiently detect earthquakes in an affordable way which will be very important in socio-economically challenged developing countries.

## VI. CONCLUSION AND FUTURE WORK

The goal is to continue the project by finding partners to implement the system on a country-by-country basis. A GoFundMe fundraiser has been created with the hope of raising funds for the first 50 seismic detection stations in the Philippines.